\documentclass[runningheads]{llncs}
\usepackage[numbers,sort&compress]{natbib}% Citation support using natbib.sty
\usepackage[utf8]{inputenc}
\usepackage{fancyhdr}
\usepackage{mathtools}
\usepackage{tikz}
\usepackage[hidelinks]{hyperref}
\usepackage{graphicx}
\usepackage[
lambda,
advantage,
operators,
sets,
adversary,
landau,
probability,
notions,
logic,
ff,
mm,
primitives,
events,
complexity,
asymptotics,
keys
]{cryptocode}
\usepackage{float}
\usepackage{authblk}
\usepackage{qtree}
\usepackage{multirow}

\providecommand{\keywords}[1]
{
	\small	
	\textbf{\textit{Keywords---}} #1
}

\DeclareMathOperator{\trap}{\pcalgostyle{IdTrapdoor}}
\DeclareMathOperator{\nulls}{\pcalgostyle{IdNullifier}}
\DeclareMathOperator{\nullss}{\pcalgostyle{Nullifier}}
\DeclareMathOperator{\extnull}{\pcalgostyle{ExtNullifier}}
\DeclareMathOperator{\secs}{\pcalgostyle{IdSecret}}

\DeclareMathOperator{\signal}{\pcalgostyle{Signal}}
\DeclareMathOperator{\proofx}{\pcalgostyle{Proof}}
\DeclareMathOperator{\data}{\pcalgostyle{Data}}

\DeclareMathOperator{\smartc}{\pcalgostyle{SC}}
\DeclareMathOperator{\commx}{\pcalgostyle{Command}}
\DeclareMathOperator{\votex}{\pcalgostyle{Vote}}
\DeclareMathOperator{\tx}{\pcalgostyle{Transaction}}
\DeclareMathOperator{\encx}{\pcalgostyle{Enc}}
\DeclareMathOperator{\sharekey}{\pcalgostyle{SharedKey}}

\newtheorem{Prop}{Proposition}

\newif\ifFIXESON
\FIXESONtrue % coment this line to remove fixme comments
\ifFIXESON
\newcommand{\fixn}[2]{\fixfootnote{\textbf{#1:} #2}}
\usepackage[draft,inline,nomargin,index]{fixme}
\else 
\newcommand{\fixn}[2]{}
\usepackage[final]{fixme}
\fi
\fxsetup{theme=color,mode=multiuser}
\definecolor{jade}{HTML}{00A86B}
\definecolor{tan}{HTML}{D2B48C}
\FXRegisterAuthor{am}{envam}{\color{red}AM}  %Aida
\FXRegisterAuthor{ag}{envag}{\color{green}AG}   %Andrea

\fxsetup{layout={footnote}}
\fxusetheme{colorsig}
                 
\begin{document}

\title{Towards a Privacy-Preserving Dispute Resolution Protocol on Ethereum}
\author{Andrea Gangemi\inst{1}\orcidID{0000-0001-9689-8473}, Aida Manzano Kharman\inst{2}\orcidID{0000-0002-5342-3037}}
\authorrunning{A. Gangemi, A. Manzano Kharman}
\institute{Department of Mathematics, University of Trento, Povo, 38123 Trento, Italy 
\and Department of Design Engineering, Imperial College London, United Kingdom\\
\email{andrea.gangemi@unitn.it, aida.manzano-kharman17@imperial.ac.uk}}

%

%\institute{Department of Design Engineering, Imperial College London, United Kingdom\\
%\email{aida.manzano-kharman17@imperial.ac.uk}}

\date{}
\maketitle                          % typeset the header of the contribution
\begin{abstract}
We present a new dispute resolution protocol that can be built on the Ethereum blockchain. Unlike existing applications like Kleros, privacy is ensured by design through the use of the zero-knowledge protocols Semaphore and MACI (Minimal Anti-Collusion Infrastructure), which provide, among other things, resistance to Sybil-like attacks and corruption. Differently from Kleros, dispute resolution is guaranteed despite the users having the final say.  Moreover, the proposed model does not use a native token on the platform, but aims to reward stakeholders through a social incentive mechanism based on soulbound tokens, introduced by Weyl, Ohlhaver, and Buterin in 2022. Users with these tokens will be considered trustworthy and will have the ability to govern the platform. As far as we know, this is one of the first blockchain projects that seeks to introduce social governance rather than one based on economic incentives.

\end{abstract}
\emph{\keywords{Dispute Resolution, Ethereum, Quadratic Voting, Soulbound Tokens, Zero Knowledge Proof}}

\section{Introduction}\label{Sect:Intro}
Distributed Ledgers Technologies (DLTs) have been the enablers of the Web3 revolution: people now actively participate in the services they use, and they want a share of the reward. Motivated by the pursuit to decentralise power, technologies such as Blockchain began to emerge. 
These architectures were designed to enable collective decision-making. 
Removing centralisation means more people have a say. As a consequence, conflicts may arise: participants may have diverging views. 
To avoid conflict, DLTs have an underlying protocol that participants agree to follow. Namely, DLTs must have one version of the truth. However, since their proliferation, a new wave of disruptive technology has emerged: Decentralised Applications, (\textit{dApps}). These are projects or organisations built on DLTs that leverage their architecture to operate in a decentralised manner. The DLT protocol serves as the service layer, upon which applications are built to provide a service or achieve a collective goal. 
For example, dApps such as \textit{Proof of Humanity} (PoH)\cite{PoH21} aim to create a Sybil-proof list of humans, using social verification to add new users. A more traditional company like Binance \cite{Bin} , instead, allows users to buy tokens or cryptocurrencies using fiat currency.\\
Conflicts and disputes of many kinds may arise. For instance: Binance must ensure that all tradeable tokens on their site are related to valid Blockchain projects, and the Proof of Humanity protocol must only add ``real" humans to its registry, not bots or fake accounts. Indeed, if Binance were to allow the trading of scam tokens, or if PoH were to allow bots to register with the service, the consequences would be catastrophic. For this reason, third parties must be able to initiate disputes if they believe that something within the application is not working as intended.

\paragraph{\textbf{Blockchain dispute resolution mechanisms.}}
Conflict resolution in DLTs and dApps is not yet regulated by law. The dApp space is rife with disputes, some leading to significant financial damage to their participants\footnote{As an example, see the following website: \href{https://www.vice.com/amp/en/article/xgd5wq/democratic-dao-suffers-coup-new-leader-steals-everything}{DAO Coup, Vice}.}. There is no dedicated protocol to protect victims or hold wrong-doers accountable. As a result, a number of dispute resolution projects have emerged as an attempt to address this issue. Kleros is the most notable one to date \cite{KlerosW}.

%The resolution of conflicts that arise within the blockchain is not regulated by laws, and to date the entire environment is wild in this respect.\\
%Furthermore, given the features of a blockchain, it may also be a helpful tool to speed-up classic off-chain disputes. In fact, advantages would be multiple: first, the time required would be considerably shorter than the time needed to settle the dispute in court; second, unlike traditional systems, which are often vulnerable to corruption, the blockchain environment is seen as a tamper-proof system that can be particularly effective in handling proceedings leading to a ruling and its automatic enforcement.\\
%For this reason, several dApps such as Kleros \cite{KlerosW} have sprung up that specifically aim to resolve these new conflicts. 
However, all these dispute resolution services have something in common: they are based on \textit{arbitration}, a form of \textit{Alternative Dispute Resolution} (ADR), a classic method to resolve disputes outside the courts. The limitation of this procedure is that the conflict passes into the hands of third parties, arbitrators, who ultimately make a decision that the users involved in the dispute are obliged to accept. DLTs, however, were designed to avoid relying on third-parties for decision-making, as this would re-introduce centralisation. Hence, we consider solutions following this approach unsuitable. 
\textit{Mediation} is another \textit{Alternative Dispute Resolution} method, which can be considered an improvement over current existing techniques. The difference from arbitration is that the users defend their stance in a conflict to one or more mediators, who eventually propose a mediation agreement. The users can choose to reject the agreement if they do not find it satisfactory. Third parties involved in the mediation process can therefore propose, but not impose, and are thus much friendlier in the Blockchain context. The drawback is that if the users reject the agreement, the dispute is not resolved and they must resort to another process to resolve the problem.

\paragraph{\textbf{Our contribution.}}
We present a new proposal for dispute resolution on the Ethereum blockchain. The proposal is designed to be implemented on Ethereum given that it is most frequently used as a service layer for dApps due to its support of smart contracts\footnote{A smart contract is a program that will automatically execute a code once certain conditions are met. It does not require intermediaries and allows for the automation of certain tasks \cite{7467408}\cite{szabo1994}.}. However, we note that in future work this proposal may be extended to use cases beyond Ethereum.
The process is divided into two phases: firstly, whenever a conflict arises, judges may attempt to resolve the dispute. They will have the opportunity to read the reason for the conflict; then, they can vote on which user they believe is in the right using a \emph{one-person-one-vote} mechanism, as well as propose an agreement that would end the dispute. \\
In the second phase, the users will vote for their preferred judges' proposals, this time using \emph{quadratic voting}, and the number of votes that are allocated will be equal to the votes they received during the first phase. The choice of \emph{quadratic voting} is justified because using a voting system such as \emph{one-person-one-vote} would result in a consistently winning strategy for the proposal that received the most votes during the first phase.\\
During the first phase, voting judges will not be able to know who others are voting for or what they voted for, while during the second phase the users involved will not know the preferences of other participants until the end. \\

\paragraph{\textbf{Structure of the paper.}}
The paper is organised as follows: in Section \ref{sect:related} we briefly recall the state-of-the-art of dispute resolution mechanisms in the blockchain environment and we survey various voting mechanisms used in the Web3 space. Section \ref{sect:preli} outlines the functioning of the key working components of the dispute resolution mechanism proposed. It illustrates \textit{Semaphore} and \textit{MACI} (Minimal Anti-Collusion Infrastructure), two zero-knowledge layers, available on the Ethereum blockchain, that we combine to enable private voting on dispute resolutions, and \emph{soulbound tokens}, used to incentivise social compliance. Furthermore, we also explain how \emph{Proof of Humanity}, a proof-of-personhood system, works, and we aim to justify why we select \emph{quadratic voting} as the voting mechanism for our proposal. All these tools will then be used in Section \ref{sect:newmec}, which introduces the framework on which our protocol is based. Later, Section \ref{sect:inc} depicts the social and financial incentives of the system, while Section \ref{sect:attacks} is devoted to attacks and describes how our protocol is able to resist them.
Finally, %Section \ref{sect:att} shows some possible attacks on the system and how it is possible to limit them, while Section \ref{sect:att} shows how it is possible to limit the system. 
Section \ref{sect:conc} outlines paths for possible future work.

\section{Related Work}\label{sect:related}
We present the survey of related work both in the context of existing conflict resolution mechanisms, as well as voting mechanisms used in Web3 dApps. Although our contribution is a dispute resolution mechanism, we require the use of a voting mechanism to enable its functioning. We explore existing voting mechanisms and later introduce the one we select and why it possesses properties that make it the most preferable choice.
\newline

\subsection{Dispute Resolution Mechanisms}
We organise the survey of existing dispute resolution mechanisms according to the following taxonomy:

\begin{itemize}
    \item \textbf{Global mechanisms}: they are Decentralised Autonomous Organisations (DAOs) that can be integrated into other DAOs and solve disputes arising from any other DApp in the ecosystem. Kleros \cite{KlerosW} is one such example.
     \item \textbf{Local mechanisms}: 
     these include dispute resolution mechanisms native to a given DAO, where the purpose of the DAO is not exclusively to enable a dispute resolution, rather, it requires one for its own functioning. Aragon \cite{AragonW} is an example of a DAO with a built-in dispute resolution mechanism.
    \item \textbf{External mechanisms}: they are traditional dispute resolution mechanisms (arbitration, mediation, and so on) that operate entirely in an off-chain court, and then notarise the result on-chain.
    \item \textbf{Enhancing mechanisms}: there are also some applications that enhance a smart contract such that it becomes legally binding \cite{OpenLaw}.
   
\end{itemize}

We focus on the first category, since our proposal is designed to be a global mechanism that is compatible with any Ethereum DAO, but we also describe how it compares to other notable dispute resolution mechanisms such as Aragon and Jur \cite{JurW} for context.

%Recently, various blockchain dispute resolution applications have been emerged. 
%The three most important ones are \textit{Kleros} \cite{lesaege2018kleros, KlerosW}, \textit{Aragon} \cite{AragonP,AragonW} and \textit{Jur} \cite{JurW}. \\
Kleros \cite{KlerosW,lesaege2018kleros} is the most active dispute resolution platform on the Ethereum blockchain. Its service is active since 2019 and it has already solved more than 1500 disputes to date. It operates as follows: provided a dispute between two users arises, one party will initiate it, and then some time is given to the other party to join and contest that dispute. Refusal to join results in automatic loss of the dispute, resulting in automatic victory for the party who initiated the conflict.  Both parties have to pay a fee in order to engage the service of a dispute resolution mechanism. 
Judges are randomly selected and then asked to vote amongst different options: the voting system implemented is the \textit{plurality voting system}. The winning option coincides with the option that receives the supra-majority of votes. At the end of the process, only the winner of the dispute gets the fee back, while the loser fee will be used to reimburse the judges for their work. %As of 2023, judges have to choose between binary options, but in the future they plan to resolve more complicated disputes with the same method. 
Examples of disputes that have been effectively resolved by Kleros include the following areas: curated lists, escrow, insurance, token listings, and some minor areas such as social networks or Gitcoin grants \cite{curatedList, escrow, insurance, token, networks, gitcoin}.\\
Aragon \cite{AragonP,AragonW} is a DAO that enables the development and maintenance of decentralised organisations running on the Ethereum Virtual Machine. The Aragon Network is an internal organisation responsible for providing the infrastructure and a range of services to support users of the platform. 
Aragon token holders can access the Aragon DAO services, and one of these services is the Aragon Court, which tries to solve disputes arising from the Aragon DAOs through a crowd-sourcing method which works exactly like the Kleros one. The Aragon Court has solved less than 50 disputes to date.\\
Finally, Jur \cite{JurW} is a legal technology company that aims to create a legal ecosystem for the management of contractual relations. They are currently undergoing a transition phase, switching from the VeChainThor blockchain \cite{vec} to Polkadot \cite{wood2016polkadot}. They have yet to solve a dispute, but are one of the most promising projects in the space. They plan to use three different arbitration protocols for dispute resolution, depending on the level of severity of the conflict in question.\\
In the past years, many other blockchain dispute resolution projects failed or were never implemented. Examples include Sagewise, Oath, Juris (for all of them, see \cite{metzger2019current}) and Aspera \cite{barbara24overcoming}. Aspera was one of the first ideas designed to provide a dispute resolution service through mediation. However, the project failed partly because of the complexity of the design, which was largely based on machine learning and artificial intelligence, and partly because mediation does not yet seem to be necessary on the blockchain, as is confirmed by the success of Kleros, in which most disputes are handled through the use of a simpler arbitration service.

\paragraph{\textbf{The Schelling game.}}
All the active services surveyed above are \textit{crowd-sourced} arbitration protocols: these protocols involve submitting the dispute to a randomly drawn group of (possibly) untrained volunteer jurors which attempt to find a fair solution in the lowest possible amount of time. They use a game-theory model known as \textit{Schelling game} \cite{schelling1980strategy}. The key contribution of this model is that a large group of independent agents with different opinions holds a collective knowledge that achieves better results in decision-making and problem-solving than individual experts. Unlike professional arbitration, which guarantees quality judgments due to the reputation of arbitrators, crowd-sourced protocols are based on a combination of game theory and crypto-economic incentives system to encourage jurors to vote fairly. As a result, the entire legal process can be modeled as a Schelling game, with the \emph{Schelling point} corresponding to the decision considered fairest by the majority. \\
Arbitrators have to deposit some tokens as stake, and if a dispute happens they can be selected to judge with a probability proportional to their amount of stake. The same arbitrator can be selected more than once, granting them greater decision power. The selected arbitrators will then vote the option they consider to be the correct one, and the option chosen by the majority of arbitrators will be the one enforced. Those who selected the winning option receive an economical reward, while the stake of those who voted otherwise will be slashed.\\
As a strength, this protocol can potentially be implemented on any blockchain with smart contracts, allowing disputes to be resolved much faster than traditional ones, and allowing anyone to vote, thus fully conveying the sense of ``freedom" and ``decentralisation" that a blockchain should offer. In addition, there is the certainty that the dispute will be resolved. However, the latter can also be considered a disadvantage, as the users must accept the decision made by Schelling's game mechanism. Two other potential problems are the lack of a system to prevent coercion attacks, i.e., judges may collude and agree to favor one party over the other, and only partial resistance to Sybil attacks.
\newline

%On Ethereum there are also many dApps that offer an arbitration service internally, in case of a dispute. Some examples are given by \hl{Magari citare alcune dApp che hanno un servizio di arbitrato interno: rivedere chat telegram con Giulio}\\
%Finally, there are also services that assign the dispute to professional arbitrators, such as \textit{OpenLaw} \cite{OpenLaw}, a project which facilitates the establishment of legal agreements on Ethereum through the drafting and implementation of smart contracts.

\subsection{Voting Mechanisms}

Next, we proceed by presenting and comparing the most commonly used voting mechanisms in Web3 dApps. In our proposal, we require the use of a voting mechanism, firstly for judges to vote for which party they believe is in the right, and secondly for each party involved to vote on the desired dispute resolution outcome. This makes it a crucial component of the proposal.
Research on voting mechanisms in the context of dApps is still young. We find that \cite{ding2023voting} and \cite{fan2023insight} propose criteria to evaluate the suitability of voting schemes, but we consider some of these criteria to be incomplete or inaccurate. \cite{fan2023insight} does not consider \emph{fairness} in their matrix to evaluate DAO voting mechanisms, and their \emph{security} notion is a heuristic with no formal mathematical or computational formalisation. The proposed classification in \cite{ding2023voting} is also inadequate. Their \emph{robustness} definition is also lacking any formalisation and their \emph{fairness} definition is only applicable if the voting protocol is a \emph{One-Person-One-Vote} (1P1V) scheme. A 1P1V scheme is used in democratic elections, assuming a functioning Sybil-protection mechanism. Given that this is harder to achieve in the context of dApps, whilst we advocate for 1P1V schemes, we also consider the fairness and security of voting schemes when voters may purchase more voting power (i.e: a \emph{One-Dollar-One-Vote} mechanism). The latter is akin to shareholder ownership models, and for some dApps, it may make more sense as a solution.  In terms of \emph{security}, we must consider formalised notions of privacy: a voter should have the right to secret ballots and should not be able to prove how they voted to anyone. This notion is formalised in the definition of \emph{Ballot Secrecy}, as defined in \cite{smyth2021ballot}. 
Without \textit{Ballot Secrecy}, voters may sell their votes, DAOs to buy votes (vote buying cartels) can arise \cite{darkDAO} and smart contracts can be written to execute and automate transactions of votes for money. Indeed, the conditions for said exchange can be verified, because the coercer can check how you voted in the absence of \textit{Ballot Secrecy}. 
We must also consider scenarios where voting power can be purchased: it must be prohibitively expensive for a malicious attacker to amass enough voting power such that they can take over the election. Finally, the voting scheme must ensure \emph{verifiability}. Voters should be able to verify that the election outcome does indeed represent the voters' votes. 
The work in \cite{2020-mind-the-gap} identifies that individual verifiability\footnote{A voter can check whether their ballot has been included.} and universal verifiability\footnote{Anyone can check that the tally of votes reflects the votes expressed in collected ballots.} plus the cast-as-intended\footnote{The cast ballot contains the vote that the voter wishes to express.} property does not yield verifiable voting systems, and that further work must be done to build a suitable security notion. This is beyond the scope of our research, thus we assume that individual and universal verifiability suffice.

We consider the following criteria:
\newline
\textbf{Fairness:} the voting mechanism does not give some voters more power than others unintendedly.\\
\textbf{Speed:} the voting mechanism can return an outcome in a timely manner.\\
\textbf{Security:} voter's votes are secret, and they cannot prove how they voted to a third party. The voting scheme used satisfies \emph{Ballot Secrecy}, as defined in \cite{smyth2021ballot}.\\

\paragraph{\textbf{Permissioned Majority Voting.}}
The most simple and commonly used voting mechanism is Permissioned Majority Voting. As a means of Sybil protection, only token holders are allowed to vote. If a proposal receives more than half the necessary votes, it is approved. It is simple to compute the outcome and easy for participants to understand. However, it suffers from an important drawback: since voting power is often acquired through tokens, a single malicious and wealthy agent may easily take over the election. It suffices for them to amass enough tokens such that with their own vote alone, they can approve their own proposals. 
Majority Voting, also known as Plurality Voting, is known to suffer of many disadvantages. Namely: groups that use this system converge to a two-party system \cite{grofman2009duverger}, it is susceptible to tactical voting \cite{dolez2017strategic}, and it may affect voter turnout \cite{kwiatkowska2020electoral}, since a winner-takes-all system does not encourage voters to vote against the most preferred option. This latter drawback would be especially exacerbated in the context of our work, given that to vote, voters must spend their tokens. In conclusion, the system is fast, but does not treat all voters equally (despite what \cite{ding2023voting} claims). 

\paragraph{\textbf{Conviction Voting.}}
In conviction voting, participants cast their vote for or against a proposal, and the longer the vote remains unchanged, the more conviction (voting power) that vote is given. Participants are allowed to change their opinion, although doing so will make their vote lose conviction \cite{ding2023voting}. Proposals are approved if they receive enough levels of conviction. This mechanism is fairer than Majority Voting, since it takes into account the wider beliefs of the community and prevents rapid acquisition of voting power. However, its major drawback is how time-consuming it is to run. 
 
\paragraph{\textbf{Token-based Quorum Voting.}}
Similarly, another method used is Token-based Quorum Voting. A minimum number of members must take part in the voting, and, if this condition is satisfied, the proposal with the most votes gets enacted. Without the quorum, the proposal will fail \cite{fan2023insight}. A number of issues arise with this approach. Participation levels can negatively impact the approval rate of proposals that get approved. In situations where the proposal is an urgent bug-fix, this can have catastrophic consequences, as the vulnerability remains exploitable until a patch is approved. Furthermore, certain agents may purposefully sabotage proposals by abstaining. Lowering the quorum level can circumvent the aforementioned issues, but this comes at the expense of security and decentralisation. 

\paragraph{\textbf{Quadratic Voting.}}
Similar to Conviction Voting, this mechanism allows voters to express the amount of conviction in their vote. Instead of amassing voting power by allowing for time to elapse, voters may acquire more than one vote and cast it. However, the cost of acquiring more votes is not linear, but rather quadratic.
The more voting power you obtain, the harder it is to acquire more of it \cite{lalley2018quadratic}. This makes it expensive to mount a heist, but it will favour malicious agents that are very wealthy. This is not a relevant concern in the context of our proposal, because voters have a limited amount of wealth they can spend on votes. This is further outlined in Section \ref{newprotocol}.
Quadratic Voting is faster to return an outcome than Conviction Voting, and fairer than Majority Voting and Token-based Quorum Voting, since it makes monopolisation of voting power prohibitively expensive. Given this compromise, we select this voting mechanism for our proposal, and further justify this decision in Section \ref{sect:preli}.

\section{Preliminaries}\label{sect:preli}
In this section, we start summarizing the privacy protocols \emph{Semaphore} and \emph{MACI} (Minimal Anti-Collusion Infrastructure) available on the Ethereum blockchain that will be used in Section \ref{sect:newmec}. These protocols use \emph{zero knowledge Succinct Non-interactive ARgument of Knowledge}, (\emph{zk-SNARKs}), a family of cryptographic algorithms that is particularly suited for the blockchain environment. Shortly, zk-SNARKs allow to  prove possession of some information, without revealing that information.\\
Then, we outline the functioning of \emph{Proof of Humanity} and how it can be integrated with Semaphore. Subsequently, we introduce \emph{quadratic voting}, and justify its use over other existing voting mechanisms. Finally, we summarise how \emph{Soulbound tokens} are used in the proposed mechanism.

\subsection{Zero knowledge protocols}
Ethereum currently has more than twenty open projects that use zero-knowledge techniques to enhance the privacy or the scalability of underlying protocols \cite{pse}. Two of these projects, Semaphore and MACI, are useful building blocks for the idea proposed in this paper. We remark that both protocols are already existing and used by various projects in the Ethereum ecosystem.

\paragraph{\textbf{Semaphore.}} \label{sema}
\emph{Semaphore} \cite{S22} is a zero-knowledge protocol which allows Ethereum users to prove their membership in a group and send signals such as votes or endorsements without revealing their identity. More in detail, Semaphore provides three functionalities:
\begin{itemize}
    \item \emph{Creation of private identities}. A user that joins a Semaphore group receives a secret/public key pair $(\sk,\pk)$. More precisely, the secret key is a tuple of three values $\sk = (\trap, \nulls, \secs)$, where $\trap$ and $\nulls$ are generated randomly, while $\secs = H(\trap || \nulls)$ ($H$ is a hash function). The nullifier is needed to avoid users signaling more than once. The public key is instead the hash of the quantity $\secs$: $\pk = H(\secs)$. The private key $\sk$ is used to generate zero-knowledge proofs;
    \item \emph{Insertion of an identity into a group}. %A Semaphore group is an anonymity set, where users are granted a key-pair associated to their identity upon joining. 
    To be part of the same group, all the users must share a common trait. Everyone is then sure that all the members possess this trait, but they do not know the real identity of these members; %A Merkle tree is used to keep track of all the members of a group: when a new user joins a group, their public key $\pk$ becomes one leaf of the Merkle tree;
    \item \emph{Sending of anonymous signals}. Signals are signed messages which are broadcast on-chain. A signal contains the following data: a vote, a membership proof, showing that the user is a member of a Semaphore group, and the proof that the same user created both the signal and the first membership proof. For most applications, it is mandatory that every member can just signal once. For this reason, each signal contains also two additional values: a public one, called $\extnull$, which is usually the ID of the Semaphore group, and then the digest $\nullss = H(\nulls || \extnull)$. Since $\nulls$ is part of the private key of a user, if two different signals have the same value $\nullss$ it means that the same user has signaled twice. To summarise, we have that
    \[ \signal = (\data, \proofx(\pk), \proofx(\data, \proofx(\pk)),\extnull, \nullss ).\]
\end{itemize}
Semaphore can thus be regarded as a Sybil-protection mechanism: each signal sent contains certain zero-knowledge proofs, generated off-chain and validated on-chain, about the sender's membership of a certain group, as well as the validity of the signal itself.
More details about the implementation of the Semaphore circuits or their smart contracts are available on the Semaphore website \cite{S22}.
%Some protocols, like \emph{UniRep} \cite{U22}, are already using Semaphore. UniRep is a private and non-repudiable reputation protocol, where \emph{users} can receive positive and negative reputation from \emph{attesters}, and voluntarily prove that they have at least certain amount of reputation without revealing the exact amount. Moreover, users cannot refuse to receive reputation from an attester.\\

\paragraph{\textbf{MACI.}} \label{maci}
MACI \cite{MACIG} stands for \emph{Minimal Anti-Collusion Infrastructure} and it is a protocol that allows users to vote on-chain with a greatly increased collusion resistance. It was proposed by Buterin in 2019 \cite{MACIP}. All transactions on a blockchain are public, so a voter can easily show to a briber which option they voted for. MACI counters this issue by using, again, zk-SNARKs to hide how each user voted, while still allowing to know the result of the tally of the votes.\\
The MACI protocol has two different actors: \emph{users}, that is the entities that send a vote, and a single \emph{trusted coordinator}, which counts the votes and releases the final result. The coordinator uses zk-SNARKs to prove that the tally has been done correctly, using only valid votes, without revealing the vote of each user.\\
Before voting, each user, who must already possess a secret/public key pair $(\sk,\pk)$ (which can be generated when joining a Semaphore group), registers their public key $\pk$ in a smart contract $\smartc_1$. %Like on Semaphore, a Merkle tree is used to keep track of registered users. In fact, all the public keys, together with a timestamp and the number of \emph{voice credits}, are saved as a leave of this Merkle tree. Voice credits can be spent by these registered users to vote. They can vote with any address, but the transaction that expresses their vote must contain an information about the registered public key. \\
Each registered user obtains some \emph{voice credits}, which can be spent to vote on some proposal. They can vote with any address, but the transaction that expresses their point of view must contain an information about the registered public key.
Finally, each user $I$ shares also a (symmetric) key $\sharekey_{I,C}$ with the trusted coordinator $C$, which is used to encrypt and decrypt transactions. In fact, in order to vote, the user $I$ will send an encrypted transaction to some poll smart contract $\smartc_2$, containing the following data:
\[ \tx = \encx_{\sharekey_{I,C}}(\sig, \commx), \]
\[\commx = (\pk_I, \votex_{\text{option}}, \votex_{\text{amount}}).\]
$\sig$ represents the signature of the user that is sending the transaction (which is obtained using the secret key $\sk_I$), while $\votex_{\text{option}}$ is the list of projects that the user wants to vote for. Finally, $ \votex_{\text{amount}}$ is the list containing the amount of voice credits the user has allocated to each project they have decided to support. \\
Users can override their previous vote if they sign a new transaction with their secret key $\sk_I$. In this case, the coordinator will consider only the last message as valid. \\
Users can also override their public key, if they sign a new transaction that contains in the transaction data a different public key $\widetilde{\pk_I}$, while still using their secret key $\sk_I$ to sign, that is 
\[ \tx = \encx_{\sharekey_{I,C}}(\sig, \commx), \]
\[\commx = (\widetilde{\pk_I}, \votex_{\text{option}}, \votex_{\text{amount}}).\]
From then on, a transaction, to be considered valid, must contain the public key $\widetilde{\pk_I}$. This feature is known as \emph{public key switching}. After this moment,  transactions must be signed with the secret key $\widetilde{\sk_I}$. Public key switching can be used to avoid bribes, since no one except the user and the trusted coordinator knows whether or not the transaction sent will be considered valid after the decryption.\\
After the voting period, the coordinator will use a third smart contract $\smartc_3$ to keep track of all the valid votes. Then, it does the tally of the votes and publishes the results. During this process, the coordinator creates two different zk-SNARKs proofs:
\begin{itemize}
    \item the first proof is published to prove that $\smartc_3$ contains only the valid messages, without revealing all the messages;
    \item the second proof is created to show that the tally of the votes was done using only valid messages, and that their individual contribution leads to the final result.
\end{itemize}
Both proofs are verified by another smart contract $\smartc_4$, specifically built to read MACI proofs. \\
%MACI has been already successfully used by some projects like clr.fund \cite{clr} to protect the fairness of a voting process in cases the funding amounts become very large.\\
However, it must be kept in mind that MACI has one major flaw, namely, it relies all its security on a central authority. If the latter were compromised, the whole protocol would fail in its intent. One possible solution might be to use an approach derived from \emph{multi-party computation}, and thus have $N$ authorities instead of just one. In this case, the system is secure if some subset of the $N$ parties is honest, however the currently implemented MACI protocol does not offer this possibility.
% \textcolor{blue}{Another possibility might be to change the approach and use homomorphic cryptography to perform vote counting, without revealing individual votes. In this case, the trust is not completely placed on a single third party, however performing operations directly on the encrypted messages is computationally onerous, thus requiring a greater amount of time. Some recent libraries developed by Zama for the Solidity programming language allow the described operations to be performed on the Ethereum blockchain\footnote{\href{https://github.com/zama-ai/fhevm}{\textcolor{blue}{Zama fhEVM}}.}}.

\subsection{Proof of Humanity}
\emph{Proof of Humanity} \cite{PoH21} (PoH) is a decentralised proof-of-personhood solution. The aim of the protocol is to ensure that every registered account is owned by a real person and that every user just holds one account. Joining the protocol is straightforward: an interested person uploads a video of themselves, and then, to be approved, an already approved user needs to verify them. Some time needs to pass, during which that user can be challenged: this can happen if the user that is trying to register is not considered human, or if it already has an account.\\
As shown in the document \cite{zkPoH}, Proof of Humanity can be integrated together with Semaphore to solve certain privacy problems. The project, called \emph{Zero Knowledge Proof of Humanity} (zkPoH), consists of a smart contract that only allows one to register as a member of a Semaphore group if the subscriber is already registered in PoH. In this way, we are sure that each member of the Semaphore group is a real person, and they can issue signals without revealing their identity.
%When a user subscribes to zkPoH, they must send the public key of the Semaphore group $\pk$, together with the account registered in Proof of Humanity. This solution is optimal to limit Sybil-like attacks.

\subsection{Quadratic voting}  \label{sect:quad}

%\paragraph{\textbf{Quadratic voting.}} 
As we have seen in Section \ref{sect:related}, \emph{quadratic voting} \cite{lalley2018quadratic} is an alternative to  other more classic voting modes, like \emph{One-Dollar-One-Vote} (1D1V), where each user can vote as many times as they want and each vote costs one dollar, or \emph{One-Person-One-Vote} (1P1V), where each user can vote exactly once. 
%These two methods have their drawbacks: in the first case, only people who really care will vote, financing the project with a large amount of money that will also cover all the other users who are interested but not enough to pay for that service. In the second case, on the other hand, every user has equal voting power, so there is no way to express how much we care about what we are voting for.\\
%Quadratic voting is a novel method that solves both these problems in a mathematical way: in 
Shortly, in this voting model, every person can vote all the time they want, but voting $n$ times will cost $n^2$.\\ % In this way, if we vote $n$ times, the total cost of these votes will be around $\frac{n^2}{2}$ (hence, the name quadratic).\\
In certain situations, it may be interesting to also allow \emph{negative voting}, i.e. a user's vote towards a project is not intended to contribute towards that project, but to penalize it in relation to all others. The idea of negative voting has been proposed by Vitalik Buterin in the context of quadratic voting \cite{negvot}. \\
Quadratic voting is a better solution compared to 1P1V, in the case that users have a different amount of votes to allocate.

\begin{Prop}
    Suppose that two users A and B have a different quantity of votes, $V_A$ and $V_B$, to allocate to different proposals. Then, if negative voting is possible, one-dollar-one-vote mechanism has an always winning strategy for the user with the higher amount of votes, while this strategy does not exist in the case of quadratic voting, when $|V_B - V_A - 2y_2| < 2\sqrt{V_Ay_2}$ with $y_2 \leq V_B$.
\end{Prop}
\begin{proof}
Without loss of generality, suppose that $V_A > V_B$. 
\begin{itemize}
    \item \textbf{1D1V}: in the case of \emph{one-dollar-one-vote}, a winning strategy for the user $A$ is to simply go all-in to the proposal they prefer, suppose $p_1$. In fact, user $B$ can either go all in to another proposal, say $p_2$, or try to use negative voting on the proposal $p_1$ and then use their remaining votes, say $y_2$, for the proposal $p_2$.
    In the former, clearly $p_1$ is the winning proposal since $V_A > V_B$; in the latter, user $B$ wants that
  \[ V_A - y_1 < y_2, \hspace{1cm} y_1 + y_2 = V_B.\]
  However, proposal $p_2$ is the winning one if and only if $V_A < y_1 + y_2 = V_B$, which is impossible by hypothesis.
  \item \textbf{Quadratic voting}: suppose again that $A$ goes all-in on the proposal $p_1$. Clearly,  if B goes all-in on another proposal $p_2$, they will never win since $V_A > V_B$. However, by using negative voting on the proposal $p_1$, they can prevent $A$ from having a winning strategy under some circumstances. In this case, $B$ wants that
  \[ (\sqrt{V_A} - \sqrt{y_1}) < \sqrt{y_2}, \hspace{1cm} y_1 + y_2 = V_B.\]
  With some manipulations, the above equation becomes
  \[ V_A^2 - 2V_AV_B + (V_B - 2y_2)^2 < 0,\]
  and it can be shown that this holds in the range
  \[ V_A + 2y_2 - 2\sqrt{V_Ay_2} < V_B < V_A + 2y_2 + 2\sqrt{V_Ay_2}. \]
  Hence, unless $|V_B - V_A - 2y_2| > 2\sqrt{V_Ay_2}$, user $B$ is not guaranteed to lose every time if $A$ plays a simple all-in strategy.
\end{itemize}
\end{proof}
%Since quadratic voting incentives the cooperation, we have to keep in mind that the idea is vulnerable to groups that are already cooperative, or to collusion attacks. This problems can (partially) be solved, for example using MACI, or by assigning a lower weight in the quadratic mechanism formula to those users for whom it is assumed that they will have the same view of the facts.\\
Quadratic voting has already been used in practice. A notable example are the \emph{Gitcoin grants} \cite{butquad}, which uses quadratic voting to allocate funds to projects in the Ethereum ecosystem.

\subsection{Soulbound tokens}  \label{sect:sbt}
\emph{Soulbound tokens} (SBTs) have been theorised by Weyl, Ohlhaver and Buterin in 2022 \cite{weyl2022decentralized}. They are non-transferable (but possibly revocable), non fungible and publicly visible tokens that encode subjective qualities like the reputation of a user or the authenticity of a piece of art. SBTs are held into wallets, and they are public by default, but, given an application, we can achieve the ``right" degree of privacy with a mix of cryptographic protocols like zero-knowledge proofs.\\
Soulbound tokens can be attested by individuals, companies or institutions. A user may thus possess a digital identity linked to the real one, which is represented by a list of SBTs, each of which provides different information about that person. In addition, a soulbound token can also be generated by a dApp when a particular situation occurs. \\
Since these tokens are non-transferable, when a user does not follow the rules of a certain protocol it might receive one, thus showing the rest of the network their negative behavior. This type of token is therefore a useful tool to introduce social compliance into the blockchain world.
%SBTs could help mitigating, together with Semaphore, Sybil attacks,  by giving more voting powers to users with more trusted soulbound tokens. Furthermore, the dApp may continuously change governance in a programmable way, shifting the board proportionally to the distribution of these tokens.

\section{The new Dispute Resolution Protocol} \label{sect:newmec}
\label{newprotocol}
In this section, we describe in detail our proposal for a novel dispute resolution mechanism. %that can be seen as an alternative to the Schelling Game currently used in the blockchain environment. \\
%The aim is to utilise the positive aspects of Kleros and Aragon, and try to improve on the shortcomings already highlighted in Section \ref{sect:related}. 
In particular, the protocol aims to:
\begin{enumerate}
    \item allow its users to have the final say on the resolution of the conflict, i.e. they should not be forced to accept the judges' decisions;
    \item allow potential judges to participate without requiring them to stake tokens to vote on a dispute resolution, as not everyone can afford it;
    \item prevent users and judges from changing their opinions after a certain time, and ensure collusion resistance for the judges;
    \item assign governance of the dApp not to those users who possess a high number of tokens, but to those who have built their reputation over time, resolving disputes and contributing to the development of the platform itself.
\end{enumerate}

The dispute resolution process can be divided into two phases (plus one subscription phase):
\begin{enumerate}
    \item \textit{Phase 0}: users that are interested in judging register in a Semaphore group upon successfully earning a Proof of Humanity;
    \item \textit{Phase 1}: once judges are notified of a dispute, they will send a transaction to a certain smart contract, containing a vote in favour of a party in the dispute, and a possible solution to the dispute. At the end of this process, the MACI coordinator computes the tally of the votes and gives a score to each user involved in the dispute. Leveraging the privacy properties of MACI, users cannot know the individual scores assigned to them by each judge;
    \item \textit{Phase 2}: the users will be able to vote for their preferred resolutions to the dispute. The votes they have received during the first phase are equal to the \emph{voice credits} each user is granted. These \emph{voice credits} are the number of votes each user can cast in favour of a dispute resolution proposal. The proposal that receives the most voice credits will be enforced.
\end{enumerate}

\subsection{Phase 0: judges' registration}
To participate, judges must be registered on both Proof of Humanity and Semaphore. This guarantees that each user has a real identity. If the registration is successful, each judge $i$ will receive a secret/public key pair ($\sk_i,\pk_i$). This public key is then also registered on the MACI smart contract $\smartc_1$: this step is essential, as we need the MACI protocol to ensure the judges do not collude. This last step also assigns to each judge one voice credit, which will be used to score the users involved in the dispute. \\
%At the same time, an external dApp that wants to this mechanism to solve the disputes must plug the dispute resolution smart contract into their contracts. We can follow the same idea introduced by Kleros and described into the proposal ERC-792 \cite{erc792}: we will refer to the smart contract (or dApp) that has a dispute as $\disputable$.

\subsection{Phase 1: voting and proposals by the judges}
Suppose now that a dispute involving $n$ different users arises. For simplicity, from now on, we assume $n = 2$, but the model can be easily generalized for $n > 2$. One of these users can activate the dispute resolution mechanism by sending to the smart contract a transaction containing an ETH fee $f$.\\
%That user will also have to list some \textit{tags} indicating the reason for the dispute. 
The other party will have to send a transaction with the same fee $f$ to join the dispute: refusing to join makes the initiating party the winner automatically. Note that there is no incentive to start a dispute fraudulently, as in this case the other party can simply join the conflict and the judges will choose them as the winner, penalising the misbehaviour of the party that started the conflict.\\
If both sides join the dispute, they must provide evidence to support their case. Since storing documents on the blockchain is an expensive operation, one solution can be to store data off-chain, saving within the blockchain only the hash of this data. Storing data off-chain is a secondary issue, which can be left to the users involved: one can choose a centralised solution, such as a cloud server, or a decentralised solution such as the \textit{InterPlanetary File System} (IPFS) protocol \cite{ipfs}, or \textit{Web3 storage} \cite{w3s}. Another alternative is the mechanism introduced by Kleros and explained in the ERC-1497 proposal \cite{erc1497}.\\
%Not all judges can participate in dispute resolution: in fact, only those who have the expertise to understand what the dispute is about can vote. This expertise is certified by SBTs: for example, to participate to a dispute whose topic is ``maths", a judge must possess one soulbound token that shows its knowledge in that field.\\
There is also the need to set two time thresholds $t_1, t_2$: judges that want to solve the dispute need to start participating before time $t_1$, and they can express their opinion until time $t_2$, which represents the end of this first phase. We also suppose that a minimum number of judges $m$ must vote, to guarantee some level of decentralisation.\\
Every judge will vote using its voice credit, using a \emph{One-Person-One-Vote} mechanism.
% Let us indicate with $T_k$ the number of affiliations (or tags) the judge $k$ has in common with the dispute tags. Let $\Sigma = \{ \sigma_1, \ldots \sigma_J \}$ be the set containing how many times each dispute tag is shared by a judge, that is each $|\sigma_j|$ counts how many judges share the tag $\sigma_j$ with the current dispute tags, and $|\Sigma|$ is equal to the number of dispute tags. Denote with $v_{i,k}$ the score assigned by the judge $k$ to the user $i$. Then, the score obtained by users $A, B$ may, for example, be equal to
% \[ V_i =  \biggl(\sum\limits_{j = 1}^{|\Sigma|} \sqrt{\sum\limits_{l = 1}^{| \sigma_j |} \frac{v_{i,l}}{T_l}} \biggl)^2 - \sum\limits_{k = 1}^m v_{i,k}\hskip0.2cm i = A,B.\]

% This weighted voting mechanism is only an example, inspired by Weyl's work \cite{weyl2022decentralized}; however, depending on the use case, it can be replaced by another voting model. The difficulty lies in finding the most appropriate model for the context.\\
When the time $t_2$ has elapsed, the MACI coordinator computes the tally of the votes: the result is saved on-chain via a commit of the tally.
The users will then be able to see the total scores $V_A$ and $V_B$ they obtained, but they will not know the individual votes they received from each judge. \\
At the end of the first phase, there are three options:
\begin{itemize}
    \item $V_A = V_B$: both users received the same amount of total votes, that is no one received an advantage for the second phase;
    \item $V_A > V_B$:  the user $A$ received more votes than the user $B$. This does not mean that $A$ won the dispute, but only that the judges expressed a preference toward that user; 
    \item $V_A < V_B$: this case is symmetrical to the previous one.
\end{itemize}

\subsection{Phase 2: users vote on the judges' proposals}
Having reached this point, the users involved in the conflict will have the possibility to read all the $m$ proposals made by the judges and vote on the ones they prefer. The voting mechanism used in this case is the quadratic voting, and the voice credits they can spend are equal to the scores $V_A, V_B$ they obtained during Phase 1: for example, user $A$ may have received 15 voice credits, while user $B$ only has 10. In this case, to avoid always-winning strategies, users can also assign a \textit{negative vote} towards the proposals they do not find appealing. \\
At the end of the process, the proposal that received the highest score is enforced. %The $\disputable$ contract will be notified, distributing the cryptocurrency immediately, if possible. Otherwise, an amount of time, e.g. 60 days, will be given to the losing party to refund the winning one. After this period, we have only two possible outcomes:
%\begin{itemize}
%    \item the winning party receives the refund. In this case, both users acted correctly and they will receive a positive reputation score.
%    \item the winning party does not receive the refund. In this case, that user will receive a negative reputation score, together with a SBT that attests the malicious behaviour. The token is linked to his account, so everyone in the Ethereum ecosystem can notice it.
%\end{itemize}
\newline
\begin{figure}[ht]
    \centering
    \includegraphics[scale=0.14]{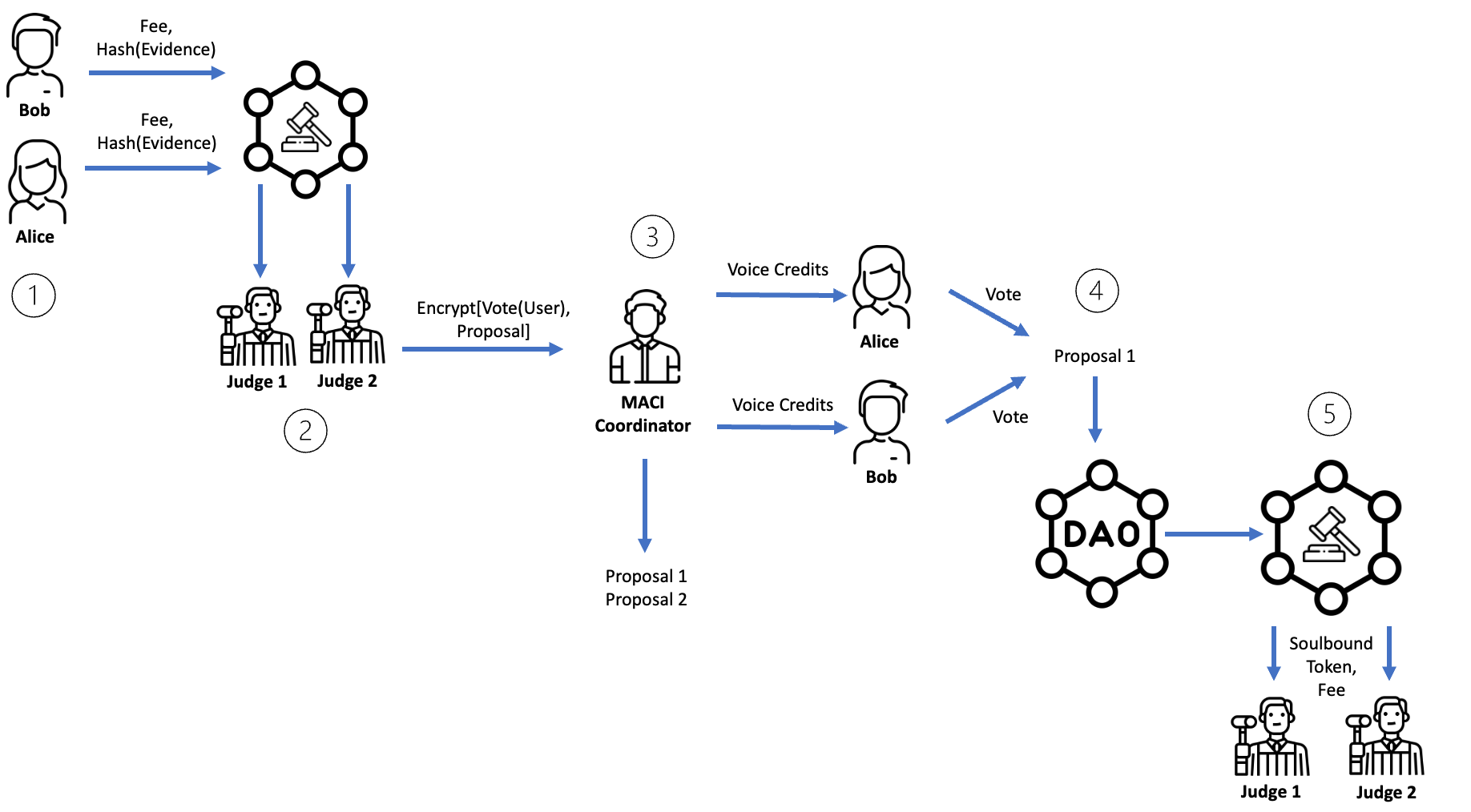}
    \caption{Dispute Resolution Mechanism.}        
    \label{fig:DRM}
\end{figure}
\newline

Figure \ref{fig:DRM} depicts the overview of the proposed solution.\footnote{Credit for icons, in order of appearance: Icons made by Freepik, Vitaly Gorbachev, LAFS, Freepik, monkik, Freepik from www.flaticon.com}  In summary, the protocol operates as follows:
\newline
1. Both conflicting parties raise a dispute to the Dispute Resolution Application (DRApp). \newline
    2. Judges that participate to the DRApp can decide if they want to rule the dispute. These vote in favour of the party they believe is right and provide a proposal for the resolution. \newline
    3. The MACI coordinator tallies the votes, which are assigned to each party as voice credits, and outputs the list of proposals. \newline
    4. Each party votes on a proposal using their voice credits. \newline
    5. The winning proposal is implemented in the DAO. The judges receive their rewards for their work through the DRApp.

\section{Incentives}\label{sect:inc}
Judges and users are incentivised to use this system in two different ways: through an economic incentive and a social incentive.\\
Social incentives reward those who behave honestly and enable the functioning of the platform, whilst penalizing bad behavior. This is possible through the use of soulbound tokens (SBTs).\\
Following Buterin’s paper \cite{weyl2022decentralized}, the ultimate goal is to entrust the DAO governance to judges who have spent their time on the proper functioning of the platform, instead of giving it to those who own more ERC-20 tokens, as is usually the case.\\

\subsection{Incentives for users}
For users who face a conflict, their main incentive is the resolution of the dispute.
Furthermore, SBTs are issued to those users who have been cooperative during the dispute process and have complied with the agreement made. They are linked to the wallet of these users, so the entire Ethereum network can see how they behaved.\\
Similarly, SBTs are issued to those users who do not comply with the agreement made. For example, suppose the dispute is about removing a token from the pool of those that can be purchased on Binance. The final proposal accepted by the users is to remove this token from the platform within a certain period of time $t$. If at the passage of this time $t$ the token is still purchasable, an SBT will be associated with the wallet of the user who was in charge of removing it. This SBT has a negative meaning, because the entire Ethereum blockchain will see that the user managing that given wallet has not fulfilled the agreement made.
%Thus, there is a social incentive for users to behave honestly in the network.

\subsection{Incentives for judges}
Unlike Kleros, judges do not need a token stake to participate to a dispute. 
However, we still need a way to incentivise good behavior and especially to penalize bad behavior, otherwise a judge can simply register and then vote effortlessly each time, effectively offering a disservice to the users. For these reasons, we need 
both an economic incentive and a social incentive. \\
The economic incentive is the fee $f$ required by the users to start the process: it will be distributed to the judge that suggested the winning proposal. All the other judges will not gain or lose anything (if we exclude the gas fee they have to pay in order to send their encrypted vote during the first phase of the protocol).\\
That is why a social incentive based on SBTs is necessary. Nevertheless, we must find a way to reward or penalize judges based on objective rather than subjective evaluations. For example, a judge who expresses a preference for the user who received fewer votes at the end of the tally has not necessarily voted in bad faith or without paying attention, and for that reason they should not be penalized. \\
However, judges should be rewarded or penalized based on the quality of dispute resolution proposals they make. During Phase 2, users vote on the proposals received using the quadratic voting mechanism. They will give a positive score if they like the proposal received, but at the same time they can also give a negative score if the proposal does not satisfy them. The score obtained from the proposals will be converted into a \emph{reputation score}, and assigned to the judges who made that proposal.\\
Next, define two thresholds $\epsilon_{-} < 0 < \epsilon_{+}$.
If a judge has a reputation score lower than $\epsilon_{-}$, they will be removed by the Semaphore group, hence they will not be able to express their opinion on the disputes anymore. Moreover, that judge will receive a SBT that certifies the bad behaviour towards the application. Notice that they cannot even create a new account and start again, since the protocol uses Proof of Humanity as a proof-of-personhood mechanism.\\
Conversely, judges with a reputation score higher than $\epsilon_{+}$ will receive a SBT which will certify their honesty and dedication: they will be considered \textit{trusted}. All these SBTs are issued by the application itself. \\
Trusted judges have the possibility to contribute towards the governance of the platform.
Thus, unlike classic dApps, everything is divided proportionally among all the users who have spent their time on the platform to make the service work. Indeed, we observe that in an application such as the one described in this paper there is no need to introduce any native ERC-20 token.

\section{Resilience to Attacks}\label{sect:attacks}
In this section we outline the attack vectors that are prevented in our novel dispute resolution mechanism.
\newline
\textbf{Voter Coercion:} Voters cannot be coerced to vote in a certain way because their vote is encrypted and an adversary cannot deduce the content of the voter's ballot. Hence, the voter has no way of proving to a coercer how they voted, and thus a coercer cannot ensure that their victim complied. 
The only agent with access to a voter's content is the MACI coordinator, who shares a symmetric key with the voter to tally the votes. It is assumed the MACI coordinator is a trusted entity, and in future work alternatives to distribute tallying may be explored. 
\newline
\textbf{Vote Selling: }
Similarly, vote selling and buying is not a relevant concern, since the buyer cannot verify the content of their purchased vote. Voters may at most sell their private key to encrypt their vote, however, in doing so they forgo their identity, and therefore their ability to participate in any other election, because their key-pair containing their secret key is generated through the Semaphore group inclusion. 
\newline
\textbf{Voter Information Asymmetry: }
Public votes create information asymmetry for voters participating in the election. The first voter to cast their vote knows no information about what will the election outcome be, conversely, the last voter to cast their vote can compute the election outcome themselves.
This can be exploited to mount \textit{`pump and dumps'}. A proposal may be put forth that can stand to increase the value of a DAO token. If votes are public, participants can observe that that proposal is set to be enacted if it is amassing votes. Participants may rush to acquire more of said token, but a voter that has a large amount of voice credits can withhold their vote until just before the election ends. They sell their tokens just before the end of the election, cast their vote against the proposal and the token value plummets. They have made a profit unlike all other participants. Our protocol resolves this by keeping votes secret, no external observer or voter can deduce the content of the votes as the election is taking place.
\newline
\textbf{Double Voting: }
A user cannot cast more than one vote as this is prevented through Semaphore. 
\newline
\textbf{Forcible take-overs:}
It is not possible for a voter to purchase large amounts of voting power in our protocol. Therefore, there is no way for an agent to amass sufficient voting power to win an election with only their vote. 
This is prevented in our system because qualified judges assign a finite amount of Voice Credits to each person in the dispute. Voters cannot purchase or otherwise acquire these Voice Credits.
\newline
\textbf{Spam protection: }
The protocol presented is resistant to spam attacks, because to start a conflict a user must deposit a fee. The challenged party cannot ignore the dispute because otherwise they automatically lose it. There is no incentive to maliciously start a fake dispute as this would cause the user to lose their fee deposit, and no incentive to avoid entering it because otherwise you're automatically the losing party, and the winning party's proposed dispute resolution is enacted.  
\newline
\textbf{Sybil protection: }
Users cannot gain an unfair advantage by creating a large number of fake identities. This is prevented through the Proof of Humanity component in the protocol that ensures a user is a real human and that they have not already registered in the platform.

\section{Conclusions and future work}\label{sect:conc}
We have proposed a novel dispute resolution mechanism built on Ethereum that provides a number of advantages over the state-of-the-art:
Firstly, the judges resolving disputes %are be selected according to their expertise, 
are verified, real individuals and the voting system they use prevents collusion amongst them, as well as maintaining secret ballots to prevent them from being coerced. This is achieved using two zero-knowledge protocols, Semaphore and MACI, which guarantee the privacy of all users involved and, in addition, offer protection against Sybil and collusion attacks, respectively.\\
Then, participants in the conflict vote on which proposal they wish to enact, and the voting mechanism does not grant more power to the wealthier, but rather those deemed the most trustworthy (by the judges). Finally, there is no incentive to initiate spam conflicts, and good behaviour both by the participants and the judges is incentivised and rewarded by the system. Failure to reward the other side at the end of the dispute, as well as judges' misbehavior, is socially penalized through the use of soulbound tokens that will be forever linked to their Ethereum wallet. In addition, judges who fall below a predetermined reputation score threshold are excluded from the system and will have no way to participate again. On the other hand, judges who exceed another predetermined threshold will have the opportunity to be part of the governance mechanism.\\
%The proposed idea could be a good starting point, provided that SBTs can be assigned to judges for ``knowledge fields" without having to use a trusted third party again. 

%We have proposed a dispute resolution mechanism built on Ethereum that allows for the enhancement of the arbitration paradigm, which is currently only implemented in a decentralised manner by Kleros, into a more general model that consists of a mix of the best qualities of arbitration and mediation.

Table \ref{comparisonTable} compares Kleros and Aragon with our proposal. As we can see, our idea uses various protocols that guarantee privacy by default, unlike the existing applications. Furthermore, our system incentivises social compliance and honest behaviour, as well as creating economic incentives for participation.

\begin{table}[htbp]
\centering
\scalebox{0.78}{
\begin{tabular}{|c|c|c|c|}
  \hline
  & Kleros & Aragon & \textbf{Our Contribution} \\
  \hline
\emph{Blockchain} & Any with smart contracts & Ethereum - Aragon framework only &  Ethereum \\ \hline
 \emph{Judging} & Requires a token stake & Requires a token stake & Requires a PoH account  \\ \hline 
\multirow{2}{*}{\emph{Privacy}} & Partial, thanks to the &   Partial, thanks to the & By default: MACI + \\ & use of commitments & use of commitments & Semaphore  \\ \hline

\multirow{2}{*}{\emph{Voting mechanism}}  & One-person-one-vote + &  One-person-one-vote + & One-person-one-vote + \\ & Schelling game & Schelling game  & Quadratic voting\\ \hline
\multirow{2}{*}{\emph{Resolution}} & Guaranteed, but judges & Guaranteed, but judges & Guaranteed, and conflicting \\ & have the last word & have the last word & parties have the last word \\ \hline 
\emph{Incentives} & Financial & Financial & Social compliance and financial \\
  \hline

\end{tabular}
}
\label{comparisonTable}
\caption{Comparison between Kleros, Aragon and our proposal.}
\end{table}

\paragraph{\textbf{Future Work.}} Further research is needed to figure out how to solve the problem of ``malicious but trusted" judges who try to exclude other trusted judges from the platform for their own interest. One possible solution could be the initiation of a dispute by the rest of the network of trusted judges toward the malicious judge. In addition, given the greater complexity of the protocol compared to the one proposed by Kleros, it is necessary to estimate the total costs of the gas used in order to understand, from a monetary point of view, when it makes sense to use the new model rather than the model based on the Schelling's game.\\
Finally, some components of our proposal are currently only available on the Ethereum blockchain, so for future developments we aim to generalise the solution to applications beyond Ethereum.

\section*{Acknowledgments}
Andrea Gangemi is member of GNSAGA of INdAM and of CrypTO, the group of Cryptography and Number Theory of Politecnico di Torino. Andrea Gangemi acknowledges support from Ripple's University Blockchain Research Initiative. 
Aida Manzano Kharman acknowledges and thanks IOTA
Foundation for the funding of her Ph.D studies.
The authors would like to thank Giacomo Corrias for his important comments and suggestions.

\bibliographystyle{plainnat} % use this to have URLs listed in References
\bibliography{local}
%\newpage
%\appendix
%\section{zk-SNARKs}  %\label{ap1}
%\input{AppendixA.tex}

\end{document}